\documentclass{article}
\begin{document}                                 
\noindent {\bf Transfer matrix method to study electromagnetic showers }
\vskip1.5cm
\noindent {\bf Ashok Razdan}

\noindent {\bf Astrophysical Sciences Division }

\noindent {\bf Bhabha Atomic Research Centre }

\noindent {\bf Trombay, Mumbai- 400085 }
\vskip 1.5cm
\noindent {\bf Abstract :}

Transfer matrix method gives information about underlying dynamics of
a multifractal. In the present studies, transfer matrix method
is  applied to multifractal properties of  a cherenkov image 
from which probablities of electromagnetic components are obtained.

\noindent{\bf Motivation:}

In last decade there have been many studies on the fractal /multifractal nature of
extensive air showers (EAS). Most of such studies have been reported for cosmic ray
studies [1,2] and to lesser extent [3,4] in $\gamma$- ray astronomy.
In cosmic ray studies [1,2] multifractal nature
of density fluctuations  has been
experimentally verified and Lipshitz-Holder exponent  distribution
of EAS has been found to be sensitive parameter to identify the
nature of individual EAS. In $\gamma$-ray astronomy,
it has been shown that cherenkov images
can be characterized using multifractal approach. It has been found that
cherenkov arrival time [5] is also multifractal in nature.

However, multifractal measures obtained for  cherenkov images or cosmic ray
densities do not give any information about the underlying dynamics.

Earlier Feigenbaum-Jensen-Procaccia (FJP) recognized this problem in chaos theory.
This led them to develop a method [6] which connects
multifractal measures with underlying dynamics using thermodynamic approach. In this
paper we explore the possibility of using FJP method to  study electromagnetic (EM) showers.
For this purpose we use information of $D_q$ versus q curve of a simulated
cherenkov image.

For a cascade like EAS, the underlying dynamics means that 
there are energy splits and probability variations of charged particles 
and photons.
So the idea is, to retrieve some information about energy splits or probabilities or both, 
from the Cherenkov image  using FJP method.
Since we are using P-model [7] approach, we will be able to
obtain information about  probabilities only because in P-model
there is assumption of equal split of  energies.

\noindent{\bf FJP Method:}

In this section we discuss physical outline of FJP method. The detailed mathematical approach
is given in the references [6,7 ]. The core of FJP method is a transfer matrix.
The elements of this transfer matrix are scaling functions of the dynamical process. The scaling
functions describe the contraction factors of each interval along each branch. The scaling
functions are obtained from the partition function. In general transfer matrix is $\infty$ x$\infty$
matrix. However, for practical applications mostly 2x2 or sometimes 3x3 matrix is used

For any tree structure, each parent  produces number of offsprings. At each level of refinement 
the number of offsprings are  increasing. For any two successive refinement levels, ratio R of
the partition functions can be obtained. It has been found that this ratio R= $\lambda(\tau)$
is the leading eigenvalue of the transfer matrix. The characteristic equations of this transfer
matrix is given as
\begin{equation}
\lambda^2 (\tau) - \lambda(\tau) Tr(T) +DET (T)=0
\end {equation}                 

where Tr and DET are trace and determinant of a matrix T respectively. 
By solving equation (1), information about  underlying dynamics can be
obtained.

Multiplicative processes  can be visualized in three ways.
In the first case, at each level of refinement there is unequal split but
equal probability. Such process is called L-model. In the second case,
at each level of refinement there is equal split with unequal probability.
This process is known as P-model. In the third case, there is unequal
split and unequal probability and process is known as LP-model. Most of
the problems have been solved using L or P model. Solutions of LP-model
have been found to be unstable.

\noindent{\bf Transfer matrix for EM showers:}

To study EM showers we consider P-model approach. In P-model re-arrangement
of probabilities, in the cascade results in a multifractal measure. P-models are
preferred over other models when there is no information or data available about
the underlying dynamics. The concept of multifractal measures was first conceived
in turbulence [8] by using P-model. 

For P-model [7] the ratio of partition functions for two successive refinements
is
\begin{equation}
\frac{\Gamma^{n+1} (q)}{\Gamma^{n}(q)}=\frac{\sum_{i=1} ^{N_n+1} (P_i ^{n+1} )^q}{\sum_{i=1} ^{N_n} (P_i ^{n} )^q} =R^{-\tau}
\end{equation}
where $P_i ^n$ is the probability in the ith-box for nth level of refinement.

The scaling function $\sigma_p$ is
\begin{equation}
\sigma_p(\epsilon_{n+1} ......\epsilon_0)= \frac{ P(\epsilon_{n+1},....\epsilon_{0}}{P(\epsilon_n,...\epsilon_0)} \delta_{\epsilon_n,\epsilon_n^{`}}........\delta_{\epsilon_1,\epsilon_1^{`}} 
\end{equation}

$P(\epsilon_n,...\epsilon_0)$=$P_i ^n$, where $\epsilon_i$ gives the location of probability on the
path of the tree
and  $\delta$ is Kronecker delta function.

The elements of 2x2 transfer matrix for EM showers are $\sigma_p(00)$, $\sigma_p(01)$, $\sigma_p(10)$ and
$\sigma_p(11)$. This is the  case of  one step memory  process. The binary digits 0 and 1 correspond
to the left (particle) and  right (photon) offspring of the parent. In the next level of refinement
there are two digits (00,01,10,11), the first digit denoting the offspring being left or right and
the second digit corresponds to parent being left or right.

For a given  cherenkov image, $\sigma_p(00)$, $\sigma_p(01)$, $\sigma_p(10)$ and $\sigma_p(11)$ are unknown.
Transfer matrix T
for a P-model can be written as
\[\left (\begin{array} {ccc}
\sigma_p (00) & \sigma_p (01) \\
\sigma_p (10) & \sigma_p (11) \end{array} \right)\]
with the condition 
\begin{equation}
\sigma_p(00)+\sigma_p(10)= 1
\end{equation}
\begin{equation}
\sigma_p(10)+\sigma_p(11)= 1
\end{equation}
$\sigma_p$'s correspond to the probability of particles and photons with 
$\sigma_p(00)\ne\sigma_p(01)$ and 
$\sigma_p(10)\ne\sigma_p(11)$, meaning unequal probabilities for particles and 
photons of the same parent. The characteristic equation of the transfer matrix is 
\begin{equation}
a^{2q} - [ \sigma_p ^{-\tau}(00)+ \sigma_p ^{-\tau}(11)] a^q +
[\sigma_p ^{-\tau}(00) \sigma_p ^{-\tau}(11)- \sigma_p ^{-\tau}(01) \sigma_p ^{-\tau}(10)]=0
\end{equation}

\noindent{ \bf Results:}

For a given cherenkov image whose $D_q$ versus q behaviour is known, $D_{-\infty}$, $D_{+\infty}$
can be calculated.
For a P-model
\begin{equation}
D_{\infty}= \frac{[log (\sigma_p(00))]}{[log( R^{-1})]}
\end{equation}
\begin{equation}
D_{-\infty}=\frac{[log (\sigma_p(11))]}{[log( R^{-1})]}
\end{equation}
and equation (6) for q=0, can be written as
\begin{equation}
1 - [ \sigma_p ^{-\tau}(00)+ \sigma_p ^{-\tau}(11)]+ 
[\sigma_p ^{-\tau}(00) \sigma_p ^{-\tau}(11)- \sigma_p ^{-\tau}(01) \sigma_p ^{-\tau}(10)]=0
\end{equation}
For a typical $\gamma$-ray initiated simulated cherenkov image corresponding to 50 TeV [3] energy,
$D_{-\infty}$=1.5,  $D_{\infty}$=0.6 and $D_0$=1.0 .
Using equations (7) and (8), we obtain the value of $\sigma_p(00)$=0.66 and $\sigma_p(11)$=0.34
and from equation (9), we  get the value of $\sigma_p(01) \sigma_p(10)$.
Using $\sigma_p (00)$,$\sigma_p (11)$ and $\sigma_p (01) \sigma_p (10)$,equation(6), can be solved
for different values of q to obtain $\tau (q)$. The resulting $\tau (q)$ versus q values can be
compared with simulated or experimental data.
Using equations (4) and (5), we have
$\sigma_p(00)$ = $ \sigma_p(10)$= 0.66 and  $\sigma_p(11)$ = $ \sigma_p(01)$= 0.34

\noindent{\bf Discussion:}

Feigenbaum et al [6] called multifractal measures as "static objects". Fractals/ multifractal 
measures are remenents of a complex underlying dynamics. The connection between the dynamics
and the resulting generalized dimensions obtained by Feigenbaum et al was indeed a breakthrough.
Chabbra et al [7] investigated FJP method in detail and found that $D_q$ versus q 
results, obtained using L-model,
P-model or LP-model may not always be unique. However, Chabbra et al [7] also concluded that
that FJP method will give accurate $D_q$ versus q results if (a) there is proper and independent choice of
ratio 'R'  (b) there may be some independent clue for choosing L-model or P-model or LP-model.
Thus these two conditions become important when $D_q$ versus q is calculated for comprative studies with
experimental or simulated data.

Feigenbaum et al [6] applied FJP method to chaos theory. Chabara et al [7] investigated FJP method
and applied it to the study  of energy dissipation in turbulence. Batumin and Sergeev [8]
applied FJP method to the study of intermittency in hadron collisions.

In EM showers it is well known that tree structure is of binary nature.  A $\gamma$-ray produces
$e^{\pm}$ 
pair which initiates particle / photon cascade. So at all levels of refinement there are only two
possibilities and ratio R=2 will not change. Again in  $\gamma$-ray initiated showers, it is also well known
that there is no loss of energy. For hadron initiated showers we cannot use P-model because at each
level of interaction there is energy loss.  

The resulting values of two probabilities $P_1$=0.66 and $P_2$=0.34 are obtained from transfer matrix
method of cherenkov images. These values are unique because using L-model we get equal probabilities
and LP model is inherently unstable. 
These values of probabilities  are very close to the results obtained from Heitler's model.
Heitler's model gives a simplified picture of EM showers. However, despite its simplicity
it predicts some important features of EM showers which include (a) the proportionality between
total number of particles and energy  (b) the relationship between shower maxima with energy.
Recently Heitler's [10] model has been extended to explain important features of hadron showers.

\noindent{\bf Conclusions:}

In this paper we have obtained probabilities of components of electromagnetic cascade from cherenkov
images by using transfer matrix method.

\end{document}